\documentclass[%
reprint,
superscriptaddress,
 amsmath,amssymb,
 aps,
]{revtex4-2}

\usepackage{graphicx}
\usepackage{bm}
\usepackage{color} 
\usepackage{hyperref}

\usepackage{fancyhdr} 
\begin{document}

\title{A dynamo simulation generating Saturn-like small magnetic dipole tilts}

\author{Rakesh Kumar Yadav}
\affiliation{Department of Earth and Planetary Sciences, Harvard University,  Cambridge, MA 02138, USA}
 \email{To whom correspondence should be addressed: rakesh\_yadav@g.harvard.edu}

\author{Hao Cao}
\affiliation{Department of Earth and Planetary Sciences, Harvard University,  Cambridge, MA 02138, USA}

\author{Jeremy Bloxham}
\affiliation{Department of Earth and Planetary Sciences, Harvard University,  Cambridge, MA 02138, USA}

\date{\today}

\begin{abstract}
{Among planetary dynamos, the magnetic field of Saturn stands out in its exceptional level of axisymmetry. One of its peculiar features is that the magnetic dipole mode is tilted with respect to the planetary rotation axis by only $\approx 0.007^{\circ}$ or less. Numerical dynamo simulations performed in this context have had great difficulty in producing such small dipole tilt angles without introducing {\textit{ad hoc}} ingredients such as a latitudinally varying heat flux pattern in the outer layers or stably stratified layers (SSL). Here we present a numerical dynamo simulation that generates a highly axisymmetric dynamo with a dipole tilt of about $\approx 0.0008^{\circ}$ on average. The model consists of a deep dynamo layer and an overlying low-conductivity layer but without any SSL. We highlight a novel mechanism where strong differential rotation generated in the atmospheric layer penetrates into the dynamo region, helping to maintain a very small magnetic dipole tilt. \\
{\bf Note: Published in Geophysical Research Letter}}
\end{abstract}

\maketitle

\section{Introduction}

Bounds on the dipole tilt of Saturn's magnetic field have only been tightened as {\em in situ} observations have been gathered by various spacecraft: Pioneer 11 \cite{acuna1980}, Voyager 1 \cite{ness1981}, and Voyager 2 \cite{ness1982}. Saturn flybys helped place an upper limit of about $1^{\circ}$ on the magnetic dipole tilt, while the Cassini Grand Finale in 2017 pushed it down to $\approx 0.007^{\circ}$ \cite{dougherty2018, cao2020}. Cowling's theorem \cite{cowling1933} in its original form states that a steady, purely axisymmetric magnetic field cannot be maintained by dynamo action in an incompressible fluid. While dynamos that are nearly axisymmetric but still exhibit a large dipole tilt with respect to the planetary spin axis can be imagined, such a case seems highly improbable given the importance of rotation in planetary dynamos. Therefore, in this context, we can assume that a perfectly axisymmetric magnetic field would require a dipole tilt angle of zero, which would be prohibited by Cowling's theorem. We note that the restrictions  of incompressibility and steadiness in Cowling's original proof have subsequently been relaxed; for example, see \cite{hide1982}. Furthermore, the theorem has been extended beyond purely axisymmetric fields: \cite{hide1982}  argue that Cowling's anti-dynamo theorem proof  (that considered neutral lines) can be extended to any field that is topologically similar to an axisymmetric field -- in the sense that meridional field lines define a distorted torus around the rotation axis.  In other work, \cite{kaiser2018} placed bounds on the permissible non-axisymmetry for Cowling's theorem.

To an external observer, a dynamo-generated magnetic field might appear to be purely axisymmetric and hence seemingly disallowed, but such a field might be significantly non-axisymmetric deeper within the electrically conducting regions of the planet. That non-axisymmetric field will be hidden from the external observer by a process spatially closer to the observer that axisymmetrizes the field. Stevenson \cite{stevenson1980, stevenson1982axi} was the first to propose such a process for Saturn: the metallic hydrogen region of Saturn's deep interior acts as a dynamo producing a non-axisymmetric field; on top of this dynamo layer, an electrically conducting stably stratified layer (SSL) may be present (due to differentiation of helium from hydrogen); if axisymmetric zonal flows are present in this layer then they will tend to filter out non-axisymmetric parts of the magnetic field for the external observer. 

The idea of magnetic field axisymmetrization through the presence of an SSL with zonal flows was further investigated and corroborated with 3D dynamo calculations in spherical geometry \cite{christensen2008}. Another study shortly after \cite{stanley2010} showed that the exact nature of the zonal flows (in particular, their equatorial symmetry) in the SSL is highly important, and there can be scenarios where zonal flows in the SSL may {\em destabilize} the dynamo and help produce non-axisymmetric magnetic fields. The most recent model in this context \cite{yan2021} uses a combination of a very thick SSL and enforced spatially inhomogeneous heat-flux pattern to reduce the dipole tilt. In the first two studies, the lowest magnetic dipole tilt reported was $\approx 1^{\circ}$, while the latest model reported a value of $\approx 0.066^{\circ}$. All of these are larger than the current upper limit of $\approx 0.007^{\circ}$ for Saturn. 

We investigate the Saturnian dynamo using direct numerical simulations and report a set of models which exhibit extremely small dipole tilt values, even smaller than the current upper limit at Saturn, and a relatively high level of axisymmetry in the magnetic field. Surprisingly, the model does not require the presence of an SSL or an imposed latitudinal heat-flux variation. Below we describe the model setup and the results. We then discuss how these simulations may help us to better understand Saturn's magnetic field.

\section{Methods}
We use the Lantz-Braginsky-Roberts (LBR) anelastic formulation \cite{braginsky1995,lantz1999} to model subsonic and compressible flows. This formulation assumes that the thermodynamic quantities are a combination of a static background (tilde) and small fluctuations (prime), i.e., $x = \tilde{x}+x'$. We use the entropy variable form where entropy contrast between boundaries drives the convection. In the anelastic approximation, under the ideal gas assumption, the density-stratified hydrostatic and adiabatic reference state is given by 
\begin{gather}
\frac{d\tilde{T}}{dr} = -\frac{\tilde{g}}{c_p}
\end{gather}
where the reference state temperature is  $\tilde{T}$, the gravity is $\tilde{g}$, and the specific heat at constant pressure is $c_p$, which is assumed constant. The equation of state is a polytrop where the background density and temperature are related by $\tilde{\rho}=\tilde{T}^m$, where $m$ is assumed to be 2. Gravity is assumed to be proportional to $r$, where $r$ is the radius. We refer the reader to \cite{jones2009} for a more detailed discussion about the anelastic equations used in the planetary deep-convection community.

The system consists of a spherical shell bounded by inner radius $R_i$ and outer radius $R_o$ and spinning about axis $\hat{z}$ with an angular velocity $\Omega$. The equations are non-dimensionalized using the shell thickness $d=R_o - R_i$ as length scale, time $d^2/\nu$ ($\nu$ is the kinematic viscosity) as time scale, and entropy contrast $\Delta s$ as entropy scale. The magnetic field is scaled by $\sqrt{\tilde{\rho_o}\mu_o\lambda_i\Omega}$, where $\tilde{\rho_o}$ is density on the outer boundary, $\mu_o$ is magnetic permeability, and $\lambda_i$ is magnetic diffusivity on the inner boundary.  The resulting evolution equations for velocity are:
\begin{eqnarray}
\nabla\cdot(\tilde{\rho}{\vec u})=0 \\
\left(\frac{\partial \vec{u}}{\partial t}+\vec{u}\cdot\vec{\nabla}\vec{u}\right)
= -\vec{\nabla}{\frac{p'}{\tilde\rho}} 
- \frac{2}{E}\hat{z}\times\vec{u}
+ \frac{Ra}{Pr}\tilde{g} \,s'\,\hat{r} \nonumber \\ 
+\frac{1}{Pm_i\,E \,\tilde{\rho}}\left(\vec{\nabla}\times \vec{B}\right)\times \vec{B}
+ \frac{1}{\tilde{\rho}} \vec{\nabla}\cdot \mathsf{S}, \label{eq:vel}   
\end{eqnarray}
where $p'$ is pressure perturbation, $s'$ is entropy perturbations, $\vec{B}$ is magnetic field, and $Pm_i$ is $\nu/\lambda_i$. The Ekman number $E$ is defined as $\nu/(\Omega d^2)$. The Rayleigh number $Ra$ is given by $\alpha_o g_o T_o d^3 \Delta s(c_p \kappa \nu)^{-1}$, where $\alpha_o$ is the thermal expansion coefficient at $R_o$, $g_o$ is gravity at outer boundary, and $\kappa$ is thermal diffusivity. Both thermal and viscous diffusion coefficients are assumed constant throughout the shell. The traceless rate-of-strain tensor $\mathsf{S}$ is defined by 
\begin{eqnarray}
S_{ij}=2\tilde{\rho}\left(e_{ij}-\frac{1}{3}\delta_{ij}\vec{\nabla}\cdot\vec{u}\right) \text{with}\,\,
e_{ij}=\frac{1}{2}\left(\frac{\partial u_{i}}{\partial x_{j}}+\frac{\partial u_{j}}{\partial x_{i}}\right), 
\end{eqnarray}
where $\delta_{ij}$ is the identity matrix. The equation for entropy perturbation $s'$ is:
\begin{eqnarray}
\tilde{\rho}\tilde{T}\left(\frac{\partial s'}{\partial t} +
\vec{u}\cdot\vec{\nabla} s' \right) =
\frac{1}{Pr}\vec{\nabla}\cdot\left(\tilde{\rho}\tilde{T}\vec{\nabla} s'\right) +  \nonumber \\ 
\frac{Pr\,Di}{Ra}\Phi_\nu +
\frac{Pr\,Di\,\lambda_{norm}}{Pm_i^2\,E\,Ra}\left(\vec{\nabla}
\times\vec{B}\right)^2, \label{eq:entropy}
\end{eqnarray}
where $\tilde{T}$ is background temperature and the viscous heating contribution is given by
\begin{eqnarray}
\Phi_{\nu}=2\tilde{\rho}\left[ e_{ij} e_{ji} - \frac{1}{3} (\vec{\nabla}\cdot\vec{u})^2 \right].
\end{eqnarray}
The dissipation number $Di$ is ${\alpha_o{g_o}d}/{c_p}$, where $\alpha_o , g_o$ are thermal expansivity and gravitational acceleration at the outer boundary.

The magnetic field is governed by
\begin{eqnarray}
\frac{\partial \vec{B}}{\partial t} = \vec{\nabla} \times \left( \vec{u}\times\vec{B}\right)-\frac{1}{Pm_i}\vec{\nabla}\times\left(\lambda_{norm}\vec{\nabla}\times\vec{B}\right) \label{eq:mag}  
\end{eqnarray}
where $\lambda_{norm}$ is local magnetic diffusivity normalized by its value at the inner boundary $R_i$. The magnetic field also follows the divergence-less condition 
\begin{eqnarray}
\vec{\nabla}\cdot \vec{B}=0.
\end{eqnarray}

The electrical conductivity follows a step-like change prescribed by a hyperbolic tangent function \cite{dietrich2018}: it remains constant until a certain radius and decreases by about $10^5$ in a small radius range, and stays constant afterwards (See Supplementary Fig.~1). The electrical currents are assumed to vanish beyond a certain radius to ensure a purely hydrodynamic behavior \cite{gastine2021}; this is achieved in the MagIC code by placing the insulating magnetic field boundary condition at a radius less than $R_o$. The exact location of the vanishing conductivity radius is a model input. At both boundaries, the velocity matches a stress-free condition, the entropy is held constant, and the magnetic field matches a potential field. 

The equations were solved using the open source MagIC code \cite{gastine2012}; available at {\tt https://github.com/magic-sph/magic}. The code is pseudo-spectral in nature and uses Legendre polynomials in the horizontal direction and Chebyshev polynomials in the radial direction to decompose the different variables solved. The code also adopts the toroidal-poloidal decomposition to ensure strict divergence-less criterion for the momentum density and the magnetic field. It utilizes the open source library SHTns \cite{shtns} to carry out the Legendre transforms. The simulation grid has 121 points in the radial direction, 640 points in the azimuthal direction and 320 points in the latitudinal direction.

\begin{figure*}
\includegraphics[width=0.9\linewidth]{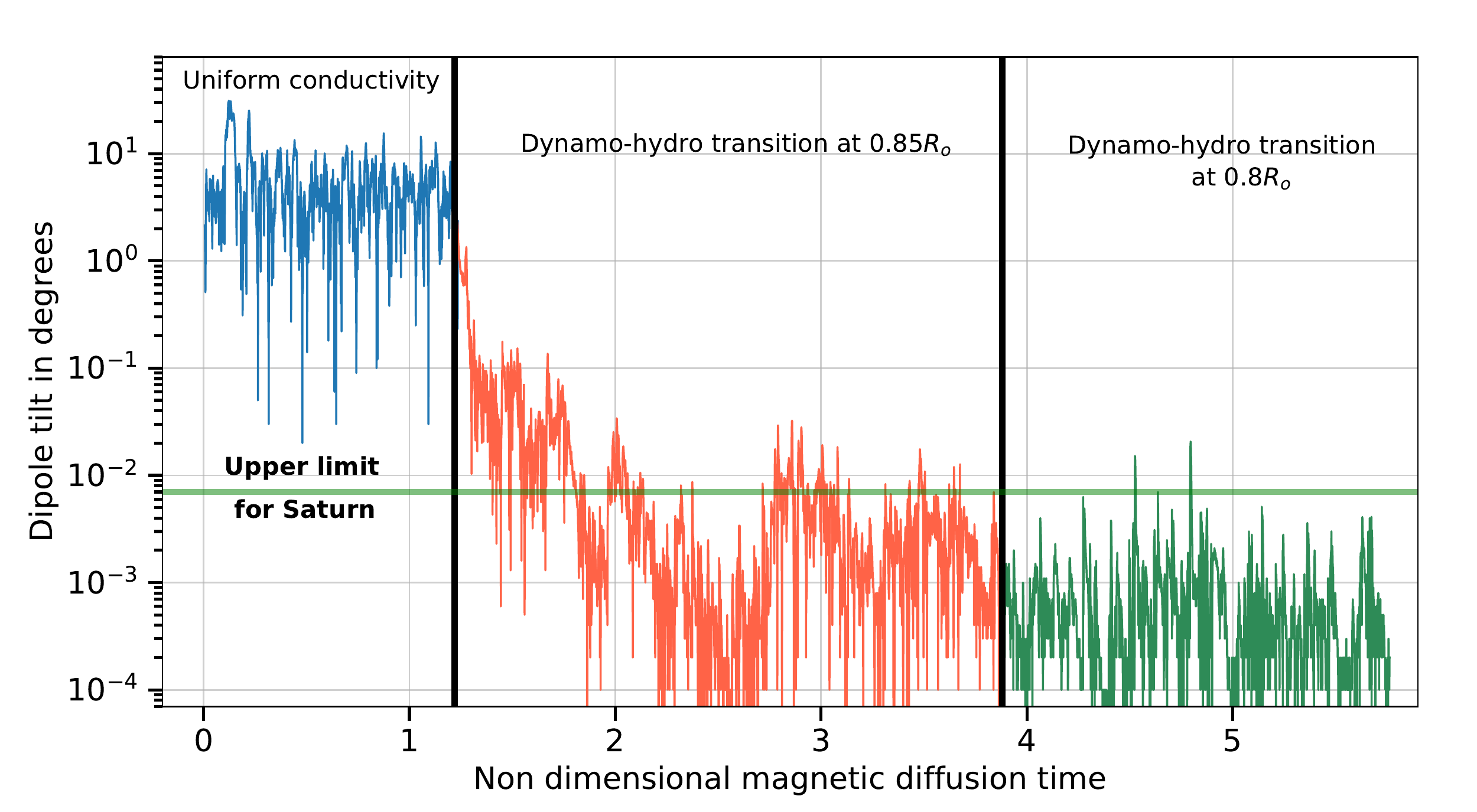}

\caption{\label{fig1}Evolution of the angle between the dipolar component of the magnetic field and the spin axis of the spherical shell as a function of the magnetic diffusion time. Here one magnetic diffusion time is defined as $d^2/\lambda_i$, where $d$ is the shell thickness and $\lambda_i$ is the magnetic diffusivity at the inner boundary $R_i$. The plot has three sections: left section shows the `baseline' case with uniform electrical conductivity fluid; the middle part shows the behavior when the electrical conductivity exponentially drops by 5 orders of magnitude at a radius around 0.85$R_o$; the right section is for a case when the conductivity drop happens around 0.8$R_o$. Note that the left most plot shows the time series only in the statistically steady state. The upper limit placed by the Cassini Grand Finale \cite{dougherty2018,cao2020} on the dipole tilt angle on Saturn's surface is shown by a thick green horizontal line.}
\end{figure*}

\section{Results}
We begin our numerical experiments by first simulating a `baseline' model that has a uniform electrical conductivity throughout its interior. The density changes by three density scale heights across the radius, translating to a change of about 20. The values of the other control parameters for this case are: Ekman number $E=10^{-5}$, Prandtl number $P_r=0.1$, magnetic Prandtl number $P_m=2$, and Rayleigh $Ra=8\times10^7$. Note that the critical Rayleigh number for convection onset for the corresponding hydrodynamic setup is about $2\times10^7$. Therefore, this dynamo setup has a low convective supercriticality of about 4. This model generates dipole tilt values $\sim 6^\circ$ (Fig.~\ref{fig1}), similar to the dipolar-branch dynamo solutions generally reported in the literature (e.g.~see \cite{duarte2018}).

We take a typical state from the baseline simulation as the initial condition and then allow the electrical conductivity to vary with radius, as is appropriate for gas giant planets \cite{stevenson1982, french2012}. We impose a rapid change in the electrical conductivity of the fluid using a hyperbolic tangent function at a certain radius. In this particular case, the step change occurs at around 0.85$R_o$ where the electrical conductivity decreases by 5 orders of magnitude within a radial thickness of about 0.1$R_o$ (a few cases with thicker transition layer produce similar results). The radial levels above 0.93$R_o$ are treated as purely hydrodynamic by enforcing a zero electrical currents condition. In this way, a deep dynamo layer, a thin layer with much smaller fluid-magnetic field interaction (which we colloquially refer to as the semi-conducting layer), and a topmost layer which has no interaction with the magnetic field are treated self-consistently. The introduction of the semi-conducting layer causes the dipole tilt angle to drop precipitously by almost four order of magnitude during a period of about 1.5 magnetic diffusion time (Fig.~\ref{fig1}). After this initial drop, the tilt angle rises slightly but stays well below the upper limit for Saturn during the majority of the simulation time. By putting the electrical conductivity drop-off location even deeper near 0.8$R_o$ and starting the zero electrical currents condition at 0.9$R_o$ (Supplementary Fig.~1) we were able to stabilize the dipole tilt angle around an average value of $\approx 0.0008^{\circ}$ (Fig.~\ref{fig1}, right-most section). 

\begin{figure}
\includegraphics[width=1\linewidth]{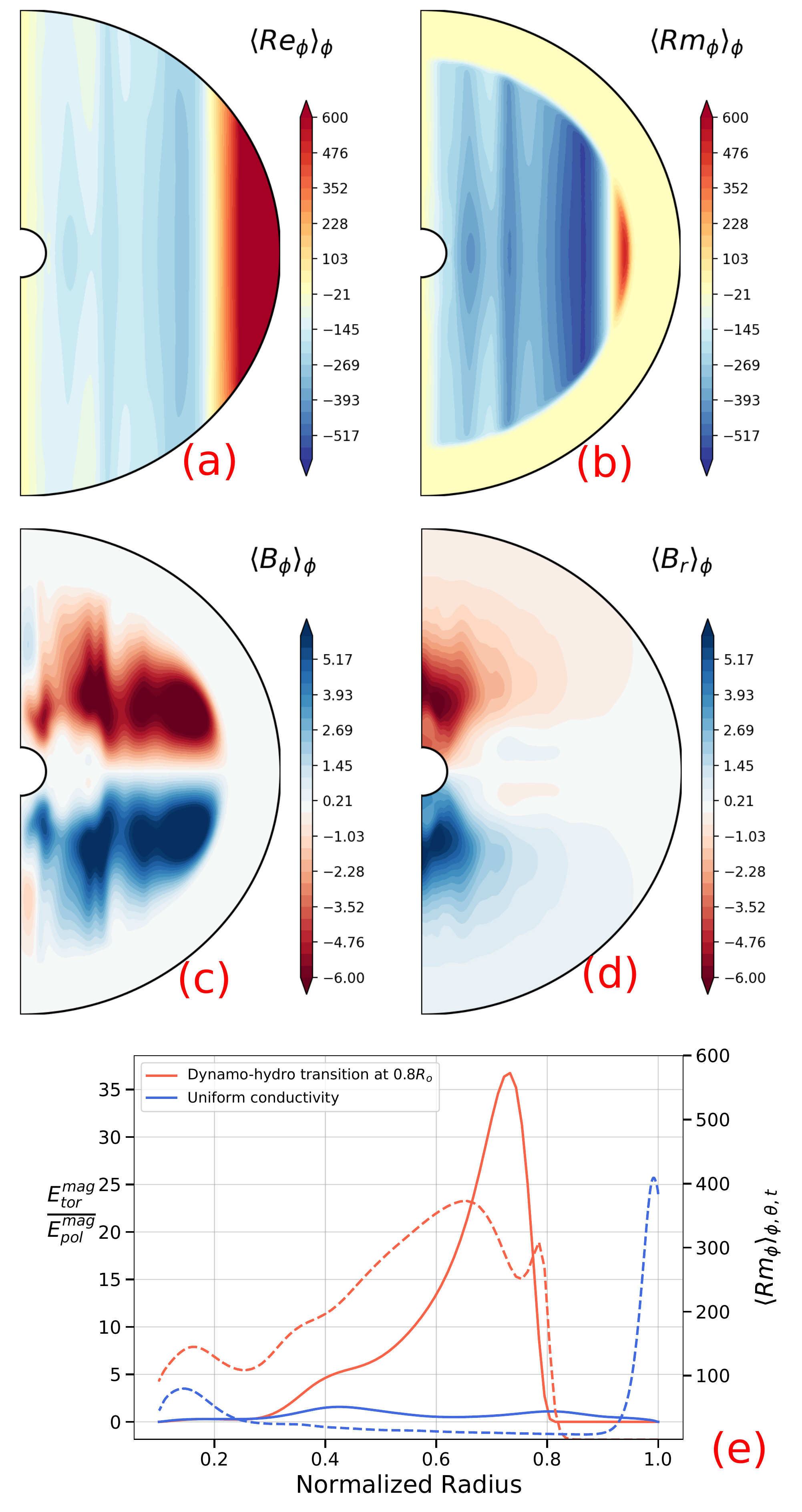}
\caption{\label{fig2}The top four panels show various azimuthally averaged (denoted by angle brackets $\langle...\rangle_{\phi}$) quantities on a meridional plane for the dynamo case with dynamo-hydro transition occurring near 0.8$R_o$: {\bf a}, azimuthal Reynolds number $Re_{\phi}$ defined as $u_{\phi}d/\nu$, where $u_{\phi}$ is azimuthal velocity, $d$ is shell thickness, and $\nu$ is viscosity; {\bf b}, azimuthal magnetic Reynolds number $Rm_{\phi}$ defined by $u_{\phi}d/\lambda$ where $\lambda$ is the local magnetic diffusivity; {\bf c}, azimuthal magnetic field given in terms of the Elsasser number units defined as $\sqrt{\Omega\mu_o\lambda_i\rho_o}$, where $\mu_o$ is magnetic permeability, $\lambda_i$ is magnetic diffusivity at inner boundary, and $\rho_o$ is density at the outer boundary; {\bf d}, radial magnetic field in the same unit. The last panel {\bf e} shows the ratio of horizontal and time averaged axisymmetric toroidal magnetic energy and poloidal magnetic energy (solid curves) on the left axis and the azimuthally, latitudinally, and time averaged $Rm_{\phi}$ (dashed curves) on the right axis as a function of normalized radius. The blue curves are for uniform conductivity simulation while the red curves are for the case where the electrical conductivity exponentially drops at around 0.8$R_o$.}
\end{figure}

The decrease in the dipole tilt angle after the introduction of a semi-conducting layer is rather remarkable. By comparing this simulation to the baseline dynamo model we can extract the main ingredients that promote low dipole tilts. As we describe below, production of a stronger zonal flow in the dynamo region and the presence of equatorially symmetric convection appears to be the key for sustaining such small dipole tilts. 

As shown in several earlier studies \cite{heimpel2011, duarte2013, gastine2014, dietrich2018, yadav2020}, the axisymmetric zonal flow increases rapidly in the hydrodynamic layer (Fig.~\ref{fig2}a) where the conductivity is low and the Lorentz forces are negligible. The magnetic Reynolds number associated with this zonal flow is appreciable only in the inner 80\% of the shell (Fig.~\ref{fig2}b, \ref{fig2}e). The profiles of the axisymmetric azimuthal (Fig.~\ref{fig2}c) and radial (Fig.~\ref{fig2}d) magnetic field show that the dynamo produces much more toroidal magnetic field compared with the poloidal component, in particular near the dynamo surface. The horizontally averaged ratio of axisymmetric toroidal and poloidal magnetic field is very high in the radius range 0.4$R_o$ to 0.75$R_o$ and reaches a maximum of around 30 (Fig.~\ref{fig2}e). The azimuthal magnetic field on the simulation surface, however, is only a few percent of the radial magnetic field (Supplementary Fig.~2). The situation is completely different in the uniform conductivity case where the toroidal and poloidal magnetic fields are roughly equi-partitioned throughout the entire volume. 

We note here that despite the larger zonal flows in the outer low-latitude regions of the dynamo-hydro setup, this setup is able to maintain a dipole-dominant state, unlike earlier studies where stronger zonal flows generally lead to a multipolar dynamo state \cite{gomez2010, heimpel2011, duarte2013, dietrich2018}. It is likely that the combination of the control parameters we use here allow the system to maintain a delicate balance between the Lorentz forces associated with the dipole dominant solution and the destabilizing effects of the zonal flows.

Such a dominance of toroidal magnetic field in cases with a hydrodynamic layer can be understood as a consequence of a strong zonal flow driven by Reynolds stress in the outer layer. As reported in earlier studies \cite{gomez2010, duarte2013}, strong eastward zonal flows are produced in the outer low conductivity, hydrodynamic layer (Fig.~\ref{fig2}a). Correspondingly, strong westward zonal flows are generated in the deeper parts of the spherical shell. In terms of magnitude, as shown in Fig.~\ref{fig2}e, the dynamo-hydro case produces almost two orders of magnitude larger zonal flow in the region where the toroidal magnetic field is highly dominant as compared with the uniform conductivity case. Such strong zonal flows drive an efficient $\Omega$-effect \cite{krause1980} which helps to produce a strong toroidal magnetic field from poloidal magnetic field. 

\begin{figure*}
\includegraphics[width=0.8\linewidth]{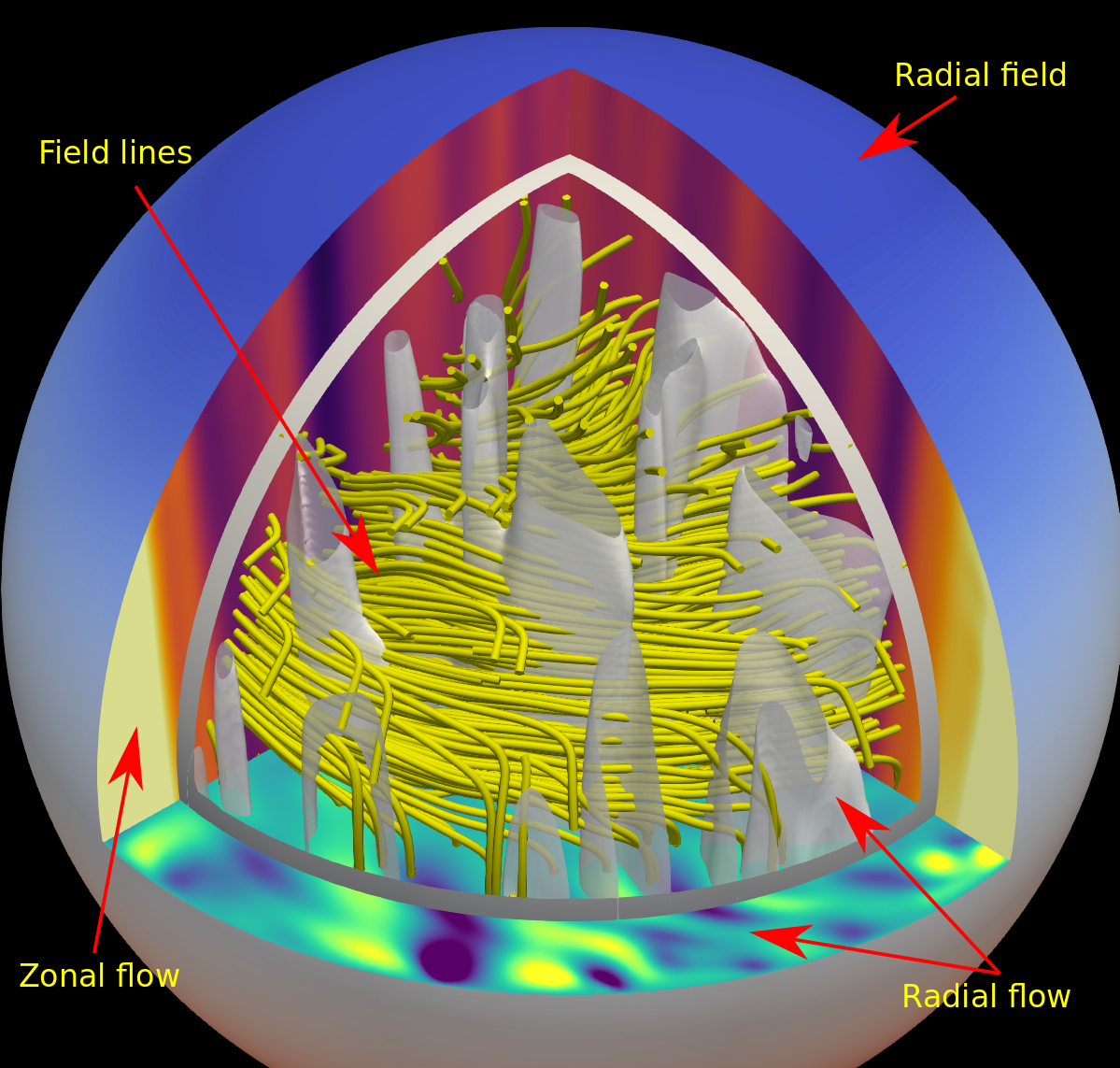}
\caption{\label{fig3}An orthographic view of the flow and magnetic field for the case where the electrical conductivity starts dropping around 0.8$R_o$. The radial magnetic field is shown by a blue-red color map on the outermost surface showing -ve and +ve values, respectively. The color map on the equatorial plane shows the radial flow, given in terms of the Reynolds number $Re$ with a range of $\pm 400$. The color map on the two meridional planes shows the zonal flow (range $\pm 1000$) at their respective locations. Here, the lighter/darker shades show +ve/-ve values. The gray spherical surface shown in the deep is one where the $Rm$ is approximately unity. The gray translucent contours embedded within the gray spherical surface are radial flow contours with an $Rm$ of 100. The magnetic field is shown using yellow tubes only in the deep interior. }
\end{figure*}

In Figure \ref{fig3} we show a three dimensional view of the magnetic field and the flow. The strong toroidal magnetic field apparent in the previous axisymmetric views is now visible in the form of magnetic field lines mainly oriented in the azimuthal direction. In the deeper region close to the rotation axis, the field lines become more poloidal, again portraying the behavior seen earlier where radial magnetic field was strongest in the innermost parts of the spherical shell where zonal flow is weaker (Fig.~\ref{fig2}d). 

The strong zonal flows are highly effective at producing strong toroidal magnetic fields that are surrounding the dipole being generated in the deepest parts of the spherical shell. The importance of zonal flows (or differential rotation) in promoting small dipole tilt values has been highlighted in the past \cite{cao2012} in a  spherical-Couette system mechanically driven by differentially rotating inner and outer boundaries. Therefore, the strong zonal flows generated in our setup are likely helping to reduce the dipole tilt values. Furthermore, a highly equatorially symmetric convection, which is excited at low convective supercriticalies (e.g., see \cite{landeau2011}), leave little room for production of magnetic field features that are different in northern and southern hemispheres. These two properties are likely pushing the dipole tilt angle to very small values. The strong zonal flows and highly equatorially symmetric convection appear to go hand-in-hand since without the hydrodynamic layer none were present in the uniform conductivity baseline model. The set-up of our simulation actively suppresses equatorially anti-symmetric convective modes since an artificial introduction of such a mode in the simulation led to a transient stage where the dipole tilt reached values as high as $1^{\circ}$ but eventually went back to the much lower values in about a magnetic diffusion time.

To investigate the stability of this small dipole tilt solution we varied some of the control parameters. Increasing the magnetic Prandtl number to 4 leads to dipole tilt values of $\approx 1^{\circ}$ and decreasing it to 1.5 leads to a no dynamo state. Increasing the Rayleigh number ($Ra=8.5\times10^7$) or decreasing it ($Ra=7.5\times10^7$) leads to much larger dipole tilt values as well. Changing the location of the conductivity drop to around 0.7$R_o$ leads to a multipolar dynamo state, while changing it to higher radius of 0.9$R_o$ leads to larger dipole tilt values. As mentioned earlier, changing the thickness of the region where the conductivity drop occurs does not disturb the dipole and maintains small tilt values. Despite some cases having stronger zonal flows, in all the instances when the dipole tilt values increased, a common theme was that the convection became less  equatorially symmetric and the ratio of toroidal to poloidal magnetic energy was significantly lower (Supplementary Fig.~3). Investigating the effects of the other control parameters, namely the Ekman number, the Prandtl number, density stratification, and the shell thickness is much more involved and requires an extensive ensemble study in future.

A recent study \cite{dietrich2018} investigating similar model setups did not report such  small dipole tilt values, although they reported an oscillatory dynamo solution with a typical dipole tile around 1$^\circ$. It used five times larger Ekman number compared to our model and generally more supercritical Rayleigh numbers. As described earlier, the dynamo state reported here is `delicate' in the sense that changing some of the control parameters leads to much larger tilt values. Modes that are symmetric or antisymmetric about the equator are excited by convection, with non-linear interactions promoting more antisymmetric ones at larger convective supercriticalities \cite{sun1995, jones2009lin, landeau2011}. One can expect that at much lower Ekman numbers where the Proudman-Taylor constraint is much stronger \cite{proudman1916, taylor1923}, convection will prefer an axially invariant flow profile, which is equatorially symmetric about the equator, in a much broader range of Rayleigh number. This would allow the dynamo model we present here to be stable in a much wider range of Rayleigh numbers. Note that Jupiter also has a semi-conducting layer on top of a deep dynamo \cite{french2012}. However, the large dipole tilt values and small scale magnetic fields observed there \cite{connerney2018,moore2018} would indicate a much more supercritical convection as compared to Saturn.

\section{Conclusions \& Discussion}
Our model here shows that dynamo solutions featuring dipole tilt even smaller than the current upper limit placed for Saturn are possible in a relatively simple setup. In other words, at least one aspect of Saturn's seemingly extraordinary magnetic field can arise from a rather ordinary model setup. The primary ingredients are axisymmetric differential rotation and equatorially symmetric convection. These ingredients were, however, present in a relatively small parameter regime at low convective supercriticalies. On Saturn, the surface zonal flow is much stronger and has high degree of equatorial symmetry. We may speculate that, if even a small fraction of the hundreds of metres/sec surface zonal flow, say a few cm/sec, reaches down to the conducting region \cite{galanti2019,kaspi2020}, it could boost the zonal flow in the deeper dynamo region making a solution of this type more likely and stable in a much wider parameter regime.  

We should note that the small dipole tilt angle is only a necessary ingredient rather than a sufficient one for describing a highly axisymmetric magnetic field such as the one on Saturn. A dynamo can have large variations in magnetic field along the azimuthal direction and still maintain extremely small dipole tilt angles as long as the field variations are equatorially antisymmetric. Our simulation produces equatorially symmetric yet non-axisymmetric magnetic field modes (Supplementary Fig.~2) which are about one order of magnitude larger as compared to Saturn (see Supplementary Tab.~1 for more details).  Furthermore, our simple dynamo model does not fully reproduce the behavior of the non-dipolar yet axisymmetric features observed at Saturn -- in particular, the drop in the magnetic energy at spherical harmonic degree 5; see Supplementary Fig.~4. We believe that the exact profile of the non-dipolar field likely is closely connected to the profile of deep zonal flows \cite{cao2017b} (see Supplementary Fig.~5). Therefore, a more complicated zonal flow profile, e.g.~one with more latitudinal variations, is likely needed to match the observed non-dipolar magnetic field at Saturn as well as help axisymmetrize the magnetic field. 

As mentioned earlier, zonal flows within a stably stratified layer are proposed as an ingredient for producing highly axisymmteric magnetic fields similar to Saturn \cite{stevenson1980, stevenson1982axi}. However, at least when it comes to small dipole tilt values, our model here shows that an SSL may not be a required ingredient. Nonetheless, recent studies have demonstrated that an SSL is likely needed on top of a dynamo layer to promote stronger zonal flows \cite{christensen2020, gastine2021} in the mid latitude regions. Therefore, a dynamical model that {\it simultaneously} explains the observed magnetic field and zonal flows still awaits construction for Saturn.

\vspace{1.5cm}

\noindent {\bf Acknowledgements}: We thank the two referees for a careful reading of the manuscript and providing constructive comments. The work was partially supported by the NASA Cassini Data Analysis Program (Grant Number 80NSSC21K1128). The computing resources were provided by the Research Computing, Faculty of Arts and Sciences, Harvard University and the NASA High-End Computing (HEC) Program through the NASA Advanced Supercomputing (NAS) Division at Ames Research Center.
\vspace{0.5cm}

\vspace{0.5cm}
\noindent {\bf Data Availability}: The software {\tt MagIC}  used to perform the simulations, as well as various analysis scripts, are freely accessible ({\tt https://magic-sph.github.io/}). Furthermore, a checkpoint file and the related input parameter file of the model with dynamo-hydro transition at $0.8R_o$ has been publicly archived on the Harvard Dataverse: {\tt https://doi.org/10.7910/DVN/AKCIZL}.

\bibliography{cited.bib}

\renewcommand\thefigure{\arabic{figure}}
\renewcommand{\figurename}{Supplementary Figure}
\setcounter{figure}{0} 

\begin{figure*}
\includegraphics[width=0.7\linewidth]{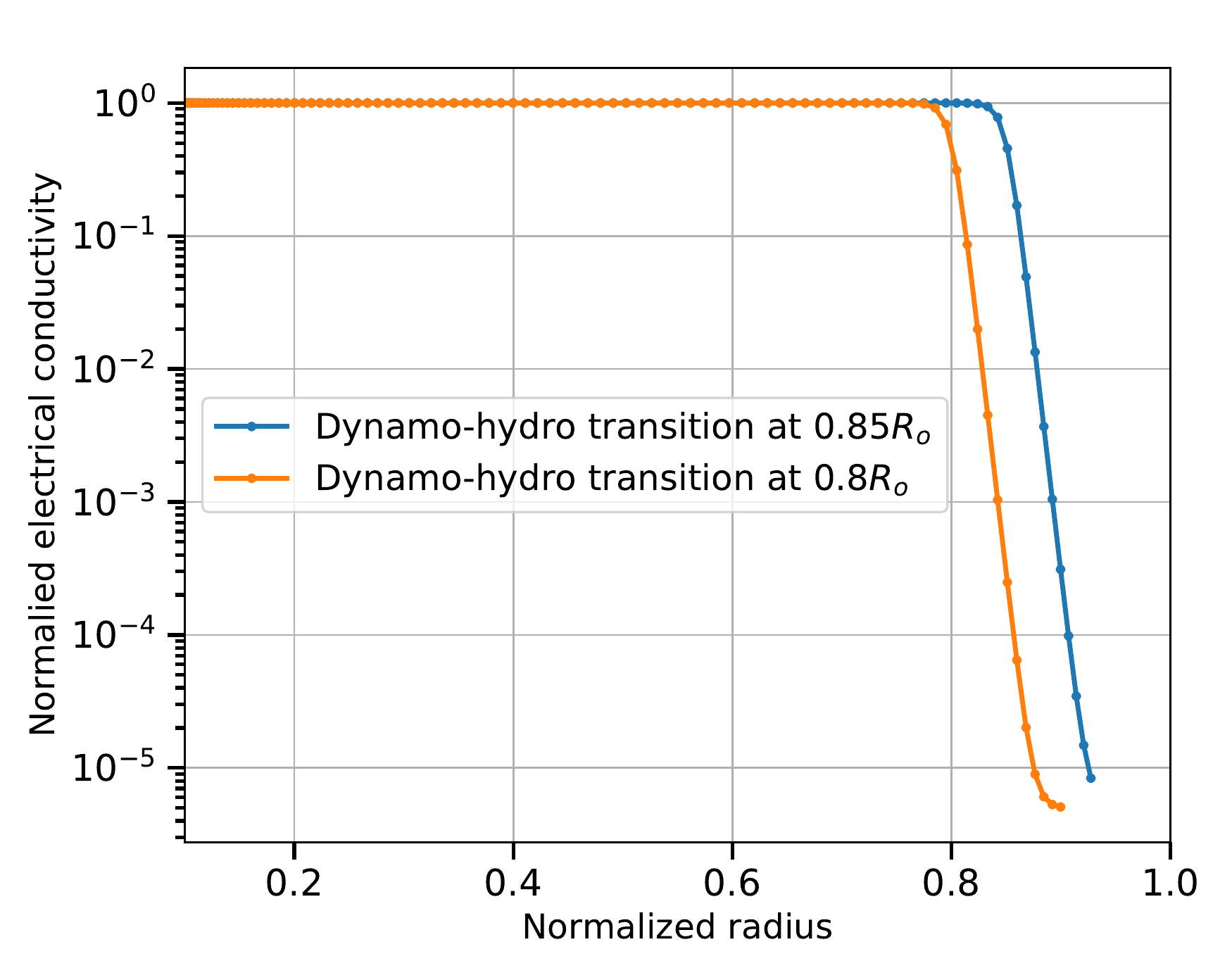}
\caption{\label{sup_fig1}Behavior of electrical conductivity as a function of radius for the two cases discussed in the paper and also shown in Fig.~1. A hyperbolic tangent function is used to define the profile; parameter option {\tt nVarCond=1} in MagIC code. Note that the profiles terminate at 0.93$R_o$ and 0.9$R_o$ since the fluid becomes completely hydrodynamic for radius above these levels.}
\end{figure*}

\begin{figure*}
\includegraphics[width=0.95\linewidth]{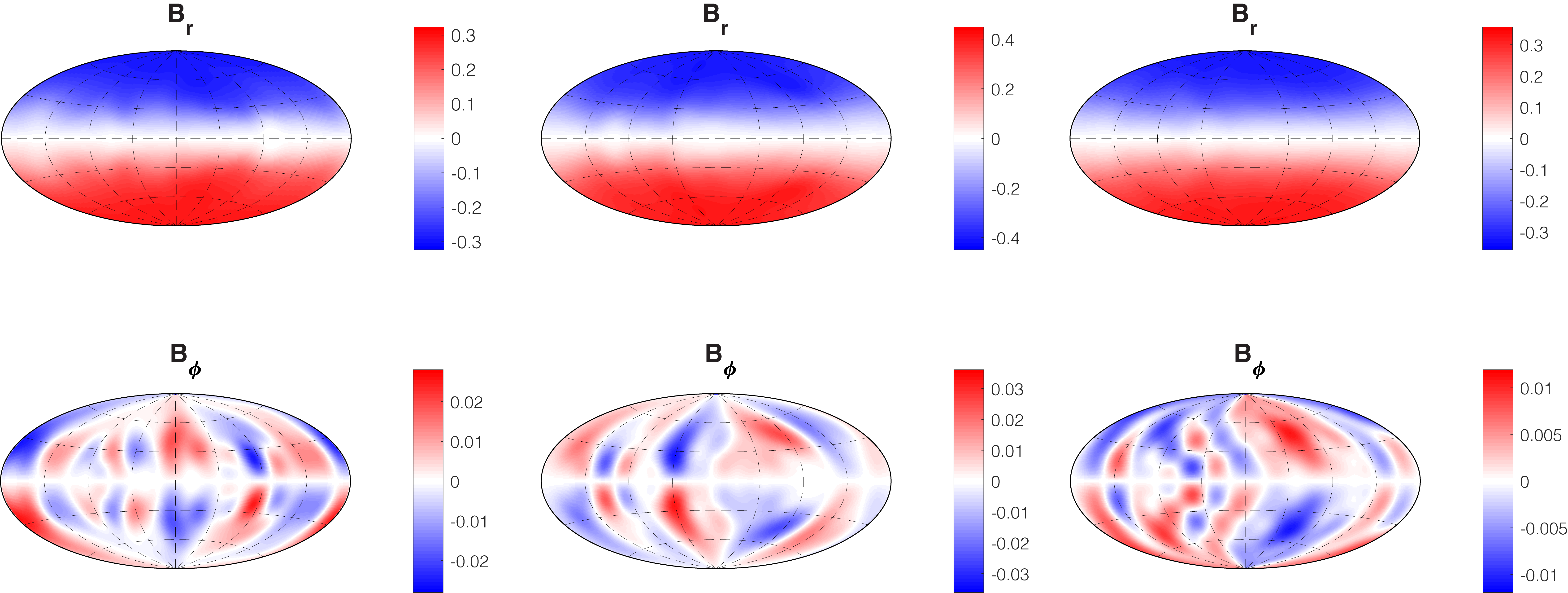}
\caption{\label{sup_fig2}The radial (top row) and azimuthal (bottom row) components of the magnetic field at the surface of the numerical simulations at three different snapshots, shown up to  harmonic degree and order 14. It can be seen that the azimuthal magnetic field $B_\phi$, which is a proxy of the total non-axisymmetry, is typically a few percent of the peak radial field.}
\end{figure*}

\begin{figure*}
\includegraphics[width=0.5\linewidth]{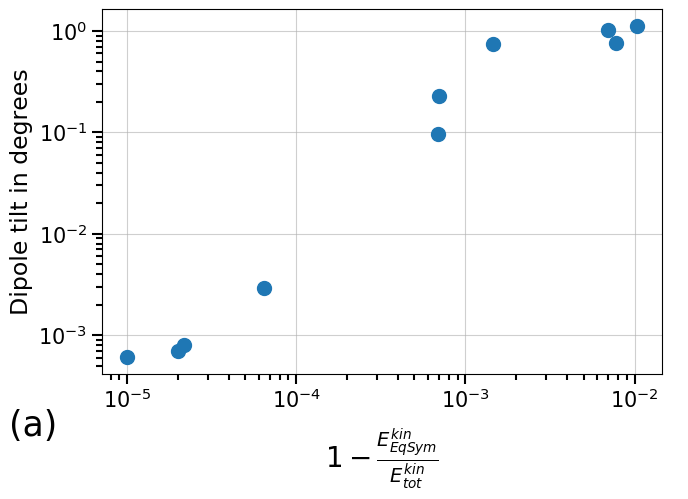} \includegraphics[width=0.5\linewidth]{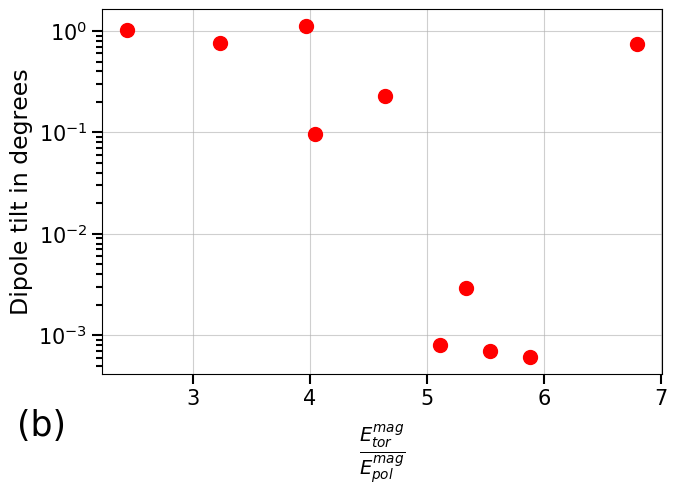}
\caption{\label{sup_fig3}The figure  shows data for several dipole-dominated dynamo simulations carried out in the study. {\bf a}, the dipole tilt versus the energy fraction in the non equatorially symmetric kinetic energy. {\bf b}, the dipole tilt versus the ratio of toroidal to poloidal magnetic energy.}
\end{figure*}

\begin{figure*}
\includegraphics[width=1\linewidth]{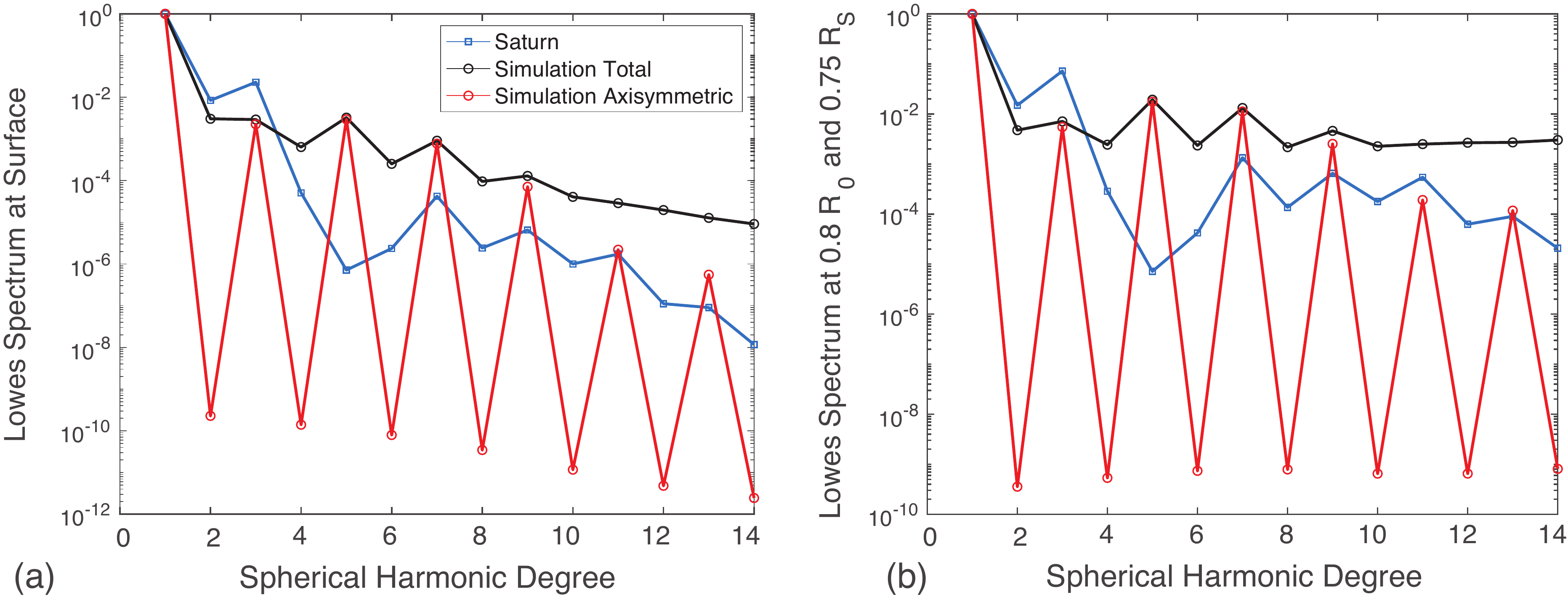}
\caption{\label{sup_fig4}Magnetic Lowes spectrum for the numerical simulation shown in Fig.~2 and 3 and for  Saturn. Panel (a) shows a comparison on the surface of the simulation and Saturn, while panel (b) shows at a deeper radial level.}
\end{figure*}

\begin{figure*}[!htb]
\includegraphics[width=0.85\linewidth]{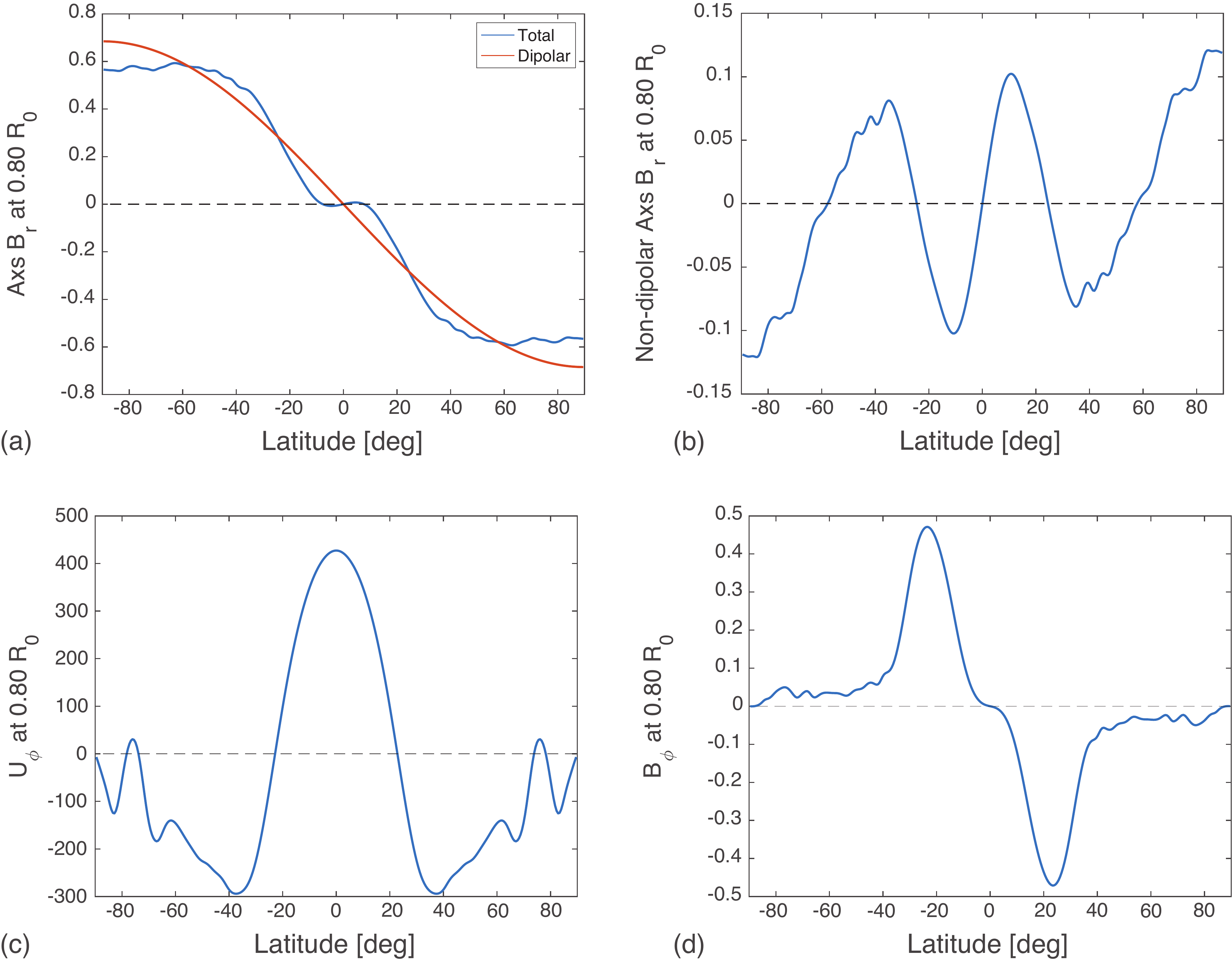}
\caption{\label{sup_fig5} Panel (a) shows the total radial magnetic field and the radial field only in the dipolar component. Panel (b) shows the radial field contained in the non-dipolar components. Panel (c) shows the zonal flow. Panel (c) shows the azimuthal magnetic field.}
\end{figure*}

\begin{table*}[!htb]
\caption{{\label{gauss_table}Normalized Gauss coefficients for the three snapshots shown in Supplementary Fig.~2. 
The last column gives the corresponding values or upper limit for Saturn based on the Cassini 11$+$ model (see Cao {\em et al.}, 2020) }.}
\centering
\begin{tabular}{r r r r r}
\hline
 & Snapshot-1 & Snapshot-2 & Snapshot-3 & Cassini 11+  \\
\hline
  $g_1^0$  & 1 & 1 & 1 & 1 \\ 
   $g_1^1$  & $-1.78 \times 10^{-8}$ & $-2.18 \times 10^{-5}$ & $-1.96 \times 10^{-5}$ & $< 1.22 \times 10^{-4}$ \\
   $h_1^1$  & $-6.51 \times 10^{-7}$ & $2.79 \times 10^{-5}$ & $-1.78 \times 10^{-6} $ & $< 1.22 \times 10^{-4}$ \\
   $g_2^0$  & $3.94 \times 10^{-6}$ & $-1.81 \times 10^{-5} $ & $-1.39 \times 10^{-5}$ & 0.0749 \\
   $g_2^1$  & -0.0013  & 0.022 & 0.018  & $<1.5 \times 10^{-3}$  \\
   $h_2^1$  & -0.028   & 0.010 & -0.011 & $<1.5 \times 10^{-3}$ \\
   $g_{2}^2$  &  $2.16 \times 10^{-6}$ & $1.69\times 10^{-6}$ &  $6.47 \times 10^{-6}$ & $<1.5 \times 10^{-3}$ \\
   $h_{2}^2$  & $-1.58 \times 10^{-6}$ & $-4.36 \times 10^{-6}$ & $3.17 \times 10^{-6}$  & $<1.5 \times 10^{-3}$\\
   $g_3^0$  & $-0.036$ & $-0.029$ & $-0.020$ & $0.107$ \\
   $g_3^1$  & $-5.56 \times 10^{-7}$ & $-9.24 \times 10^{-6}$ & $-2.9 \times 10^{-6}$ &  $<1.5 \times 10^{-3}$ \\
   $h_3^1$  & $9.20 \times 10^{-7}$ & $1.12 \times 10^{-5}$ & $8.17 \times 10^{-7}$ &  $<1.5 \times 10^{-3}$ \\
   $g_{3}^2$    & 0.0043    &  $-0.026$     & $-0.0049$ & $<1.5 \times 10^{-3}$ \\
   $h_3^2$      & $-0.0047$ &  $-0.0033$    & $-0.0051$ & $<1.5 \times 10^{-3}$ \\
   $g_{3}^3$    & $-2.21 \times 10^{-6}$ & $-1.90 \times 10^{-7}$ & $1.48 \times 10^{-6}$  & $<1.5 \times 10^{-3}$ \\
   $h_{3}^3$    & $2.28 \times 10^{-6}$ & $3.98 \times 10^{-6}$ & $-1.17 \times 10^{-6}$  & $<1.5 \times 10^{-3}$ \\
\hline
\end{tabular}
\end{table*}

\end{document}